\documentclass[cits]{PoS}

\title{Understanding the Aoki phase}

\ShortTitle{Understanding the Aoki phase}

\author{\speaker{A. Vaquero}\thanks{The authors want to thank Steve Sharpe for the fruitful discussions we held.}\\
        Universidad de Zaragoza\\
        E-mail: \email{alexv@unizar.es}}

\author{V. Azcoiti\footnotemark[2]\\
        Universidad de Zaragoza\\
        E-mail: \email{azcoiti@azcoiti.unizar.es}}

\author{G. Di Carlo\footnotemark[2]\\
        Laboratorio Nazionale di Gran Sasso\\
        E-mail: \email{gdicarlo@lngs.infn.it}}

\author{E. Follana\\
        Universidad de Zaragoza\\
        E-mail: \email{efollana@unizar.es}}

\abstract{The vacuum structure of QCD with two degenerated flavours of Wilson fermions is discussed by means of the p.d.f. (probability distribution function) formalism. Under certain assumptions, either new phases related to Aoki's appear, characterized by a non-vanishing expectation value of the condensate $i\bar\psi_u\gamma_5\psi_u + i\bar\psi_d\gamma_5\psi_d$, enriching the standard picture of the QCD vacuum structure with Wilson fermions; or the eigenvalues of the Dirac-Wilson operator must comply with a non-trivial, infinite set of sum rules, enforcing the currently accepted structure of the Aoki vacuum.\

The same scenario is analyzed using the Ginsparg-Wilson regularization. In this case, the absence of any parity and/or flavour breaking phase is proved for a non-zero value of the fermion mass.}

\FullConference{The XXVII International Symposium on Lattice Field Theory - LAT2009\\
		 July 26-31 2009\\
		 Pekine University, Beijing, China}

\begin{document}

\section{Introduction}
The understanding of the realization of symmetries in QCD from first principles has been an important issue for a long time. During the 80's, Vafa and Witten gave arguments against spontaneous breaking of parity \cite{Vafa1} and vector-like global symmetries \cite{Vafa2} in vector-like theories; however, these arguments were not as useful as expected. Some years after the publication of \cite{Vafa1}, many articles appeared \cite{agVafa} calling into question the validity of the paper (see for a recent review \cite{yo}). The fact that the issue is still open twenty years after the publication of the first paper is an indicative of the complexity of the subject. Regarding the second paper \cite{Vafa2}, it must be remarked that the result is not applicable neither to the Ginsparg-Wilson regularization\footnote{The paper \cite{Vafa2} states that vector symmetries are conserved in vector-like theories, if one is able to find an upper bound for the propagator. The paper \cite{Vafa2} fails to prove this bound for the Ginsparg-Wilson regularization, as hermiticity of the Dirac operator is used during the proof.} nor to one of the most used fermionic regularizations on the lattice for QCD, i.e., Wilson fermions. In that case, and as it was shown by Aoki \cite{Aoki,Aoki2}, there exists a region of the parameters where parity and flavour symmetries are spontaneously broken, for the conditions of the Vafa-Witten theorem are not fulfilled in the Wilson regularization. In the end, a theoretical proof of the realization of symmetries of QCD is still lacking.

This is where the p.d.f. formalism can help. The p.d.f. formalism is a powerful tool to analyze the symmetries of a theory, widely used in statistical mechanics, and introduced around ten years ago in quantum field theories for Grassmann degrees of freedom with success \cite{pdf}. In this paper, we apply the p.d.f. formalism to different regularizations of lattice QCD. The next section is devoted to a brief introduction to the p.d.f. formalism. In the second section, we analyze the Aoki phase, to find either the existence of a new phase, or a infinite set of sum-rules, the eigenvalues of the Wilson-Dirac operator must comply with. The third section applies the same formalism to another system; we successfully find, by means of the p.d.f., rigorous proof of parity and vector-like symmetries conservation in the Ginsparg-Wilson regularization of lattice QCD at non-zero mass. The last section summarizes our conclusions

\section{The p.d.f. formalism}

The usual way to study spontaneous symmetry breaking on the lattice consists in the following procedure: An external source, which breaks the analyzed symmetry explicitly, is added. This generates a non-zero expectation value of the order parameter for that symmetry. Then we take, in this order, first the thermodynamic limit, and finally, the zero external source limit. If the order parameter expectation value is non-zero after these two limits, then the symmetry is spontaneously broken. Although very popular, the method requires extrapolations to be made. Moreover, in some systems, the external source method can not be applied in lattice simulations, for the symmetry breaking term may add a potentially problematic sign problem. This is the case of the diquark condensate in two colours QCD \cite{VicSU2}. It would be desirable to be able to study the fate of the symmetries without having to add an external source.

The p.d.f. formalism enables us to do so. It simply amounts to compute the following quantity

\begin{equation}
P(c) = \lim_{V\rightarrow\infty} \left\langle\delta\left(\frac{1}{V}\sum_x\mathcal{O}(x) - c\right)\right\rangle,
\label{pdf}
\end{equation}
with $\mathcal{O}(x)$ the order parameter to be studied. In our case, this is a fermionic bilinear $\bar\psi(x)O\psi(x)$, where $O$ is a constant matrix. To obtain some computable quantity, we prefer to work with the Fourier transform of (\ref{pdf})

\begin{equation}
P(q) = \int dc\,e^{iqc} P(c) = \lim_{V\rightarrow\infty} \frac{1}{Z}\int\left[dU\right]d\psi\,d\bar\psi\,e^{-S_G}e^{\bar\psi\left(\Delta + O\frac{iq}{V}\right)\psi} =$$
$$\lim_{V\rightarrow\infty} \frac{1}{Z}\int\left[dU\right]d\psi\,d\bar\psi\,\det{\left(\Delta + O\frac{iq}{V}\right)} = \lim_{V\rightarrow\infty}\left\langle\frac{\det(\Delta+O\frac{iq}{V})}{\det\Delta}\right\rangle,
\label{pdfFT}
\end{equation}
where $\Delta$ is the Dirac operator, and $S_G$ is the pure gauge action. The expectation values of the fermionic bilinear can be computed from (\ref{pdfFT}) easily, taking derivatives of $P(q)$ at $q=0$,

\begin{equation}
\left.\frac{d^n P(q)}{dq^n}\right|_{q=0} = \left.\int dc\,(ic)^n e^{iqc} P(c)\right|_{q=0} = i^n \left\langle c^n \right\rangle.
\end{equation}
Thus, the moments of the Fourier transform of the distribution function are the expectation values of the powers of the observables. For a broken symmetry, the expectation value of the order parameter $\langle c\rangle$ will be zero, for a broken symmetry gives rise to symmetric vacua, and the expectation values of the order parameter in those vacua cancel each other. Then, the interesting observables to find broken symmetries are $\langle c^n\rangle$ with $n$ even.

\section{The Wilson scenario}

Let's apply this machinery to QCD with two degenerated flavours of Wilson fermions \cite{yo2}. In this scenario, there exists a phase -the Aoki phase- where parity and flavour are spontaneously broken, and this translates into a non-zero value of the fermionic bilinear $i\bar\psi\gamma_5\tau_3\psi$. Surprisingly, the expectation value of the bilinear $i\bar\psi\gamma_5\psi$ is equal to zero; this phenomenon was explained in \cite{Aoki2,Sharpe2}, and a brief hint will be given here: There is a $U(1)$ remnant of the original $SU(2)$ flavour symmetry, which combines with the original parity operator $P$ to yield zero expectation value of $i\bar\psi\gamma_5\psi$. In other words: We can find a redefinition of parity (a combination of parity and U(1)) which remains unbroken. This is the standard picture of the Aoki phase.

Now we compute these two fermionic bilinears using the p.d.f. formalism:

\begin{equation}
\langle\left(i\bar\psi\gamma_5\psi\right)^2\rangle = 2\left\langle\frac{1}{V^2}\sum_{j}\frac{1}{\lambda_j^2}\right\rangle - 4\left\langle\left(\frac{1}{V}\sum_{j}\frac{1}{\lambda_j}\right)^2\right\rangle,
\label{Parity}
\end{equation}

\begin{equation}
\langle\left(i\bar\psi\gamma_5\tau_3\psi\right)^2\rangle = 2\left\langle\frac{1}{V^2}\sum_{j}\frac{1}{\lambda_j^2}\right\rangle,
\label{Flavour}
\end{equation}
where $\lambda_j$ the eigenvalues of the $\gamma_5\Delta$ operator at zero external source. The expression (\ref{Parity}) can be easily generalized to any number of flavours.

We must remark that the p.d.f. can not predict what values will the expressions (\ref{Parity}) and (\ref{Flavour}) take. These depend on the specific properties of the eigenvalues of the chosen discretization of the Dirac operator. In fact, as we will see later, the results change dramatically for the Ginsparg-Wilson regularization, even though the expressions (\ref{Parity}) and (\ref{Flavour}) remain the same.

For (\ref{Parity}) to be zero, as the current picture of the Aoki phase demands, the following cancellation must happen

\begin{equation}
2\left\langle\frac{1}{V^2}\sum_{j}\frac{1}{\lambda_j^2}\right\rangle = 4\left\langle\left(\frac{1}{V}\sum_{j}\frac{1}{\lambda_j}\right)^2\right\rangle,
\label{Miracle}
\end{equation}
as we know that the left hand side of the equation must be non-zero, by virtue of (\ref{Flavour}). This relationship among the eigenvalues of the $\gamma_5\Delta$ operator is non-trivial; in fact, for every even moment of the distribution function we obtain a different sum-rule, by enforcing $\langle\left(i\bar\psi\gamma_5\psi\right)^{2n}\rangle = 0$.

From this point on, we face two different possibilities:

\begin {enumerate}
\item The standard picture of the Aoki phase is right, and these sum-rules must be fulfilled by the eigenvalues of $\gamma_5\Delta$. In fact, the p.d.f. can be used to derive easily the sum-rules, which remind to those obtained by Leutwyler and Smilga in the continuum in \cite{LeutSmilg}. This point was exposed by S. Sharpe in \cite{Sharpe}.
\item The current understanding of the Aoki phase is incomplete, for it seems improbable that the eigenvalues of $\gamma_5\Delta$ comply with $\langle\left(i\bar\psi\gamma_5\psi\right)^{2n}\rangle = 0$ for any value of $n$. So there must exist a new phase which verifies $\langle\left(i\bar\psi\gamma_5\psi\right)^{2n}\rangle \neq 0$ for some value of $n$. As $\chi PT$ predicts the standard picture for the Aoki phase, the realization of this case would imply that $\chi PT$ is, in some sense, incomplete. An analysis of this point of view was done in \cite{yo2}.
\end{enumerate}

\noindent At this moment, there is no theoretical proof to decide between one of this two realizations. In order to distinguish which one occurs, a dynamical fermion simulation in the Aoki phase is mandatory, measuring the eigenvalues of the $\gamma_5\Delta$ operator, and computing the sum-rules.

\section{The Ginsparg-Wilson scenario}

As we have seen, the original Vafa and Witten theorems fail, due to the existence of exceptional configurations, in the later scenario, which, on the other hand, is of paramount importance in lattice QCD. So, is there any way we can say something concrete about QCD symmetries?. The answer is yes, but we need to choose a `small eigenvalue free' regularization. It happens that the Ginsparg-Wilson\footnote{Strictly speaking, we will work with a Dirac operator which satisfies $\left\{D^{-1},\gamma_5\right\} = a\gamma_5$, whereas the Ginsparg-Wilson regularization only requires $\left\{D^{-1},\gamma_5\right\} = aR\gamma_5$, with $R$ a local operator. Nevertheless, the results can be applied to any version of Ginsparg-Wilson fermions, although the calculations may eventually become harder.} fermions fulfill this requirement; the p.d.f. will do the rest.

We denote by $D$ the Ginsparg-Wilson operator; as we know, its eigenvalues $\bar\lambda_j$ are complex, and lie in a circumference of radius $\frac{1}{a}$ in the complex plane, whose center is in the real axis, at the point $\frac{1}{a}$. Using the standard properties of the Ginsparg-Wilson operator, we can compute the eigenvalues of the hermitian operator $\gamma_5\left(D+m\right)$, which are

\begin{equation}
\mu_{\pm j} = \left\{\begin{array}{ll}
\pm\sqrt{\left(1+am\right)\left|\bar\lambda_j\right|^2 + m^2} & \quad \bar\lambda_j \notin \mathbb{R} \\
+\textit{ and/or }-\left(\bar\lambda_j + m\right) & \quad \bar\lambda_j \in \mathbb{R} \\
\end{array}\right. ,
\label{ComplEigen}
\end{equation}
since when $\bar\lambda_j \notin \mathbb{R}$, the eigenvalues associated to $\bar\lambda_j$ and $\bar\lambda_j^*$ are paired ($\pm$), but if $\bar\lambda_j \in \mathbb{R}$, this needn't be he case, giving rise to zero modes, and non-vanishing topological charge. As an interesting and useful remark, we see that the $\mu$'s are real, and non-zero for a non-vanishing mass $m$. If we look at the expression (\ref{ComplEigen}), we can see that the modulus of these eigenvalues is bounded from below by $m$ (this was remarked in \cite{Niedermayer}). So we establish that $\frac{1}{\mu_{\pm j}^2} \leq  \frac{1}{m^2}$, which in turn implies that, for a non-zero mass, the following expectation value\footnote{If the expressions (\ref{Flavour}) and (\ref{Flavour-Dies}), or (\ref{Parity}) and (\ref{Par-Dies}) are compared, one will notice that they are identical, even though they correspont to different lattice regularizations. These expressions are regularization independent (see (\ref{pdfFT})). But the specific properties of the eigenvalues do depend on the regularization, and make the Wilson and Ginsparg-Wilson fermions behave in different ways.} is zero in the thermodynamic limit

\begin{equation}
\langle\left(i\bar\psi\gamma_5\tau_3\psi\right)^2\rangle = 2\left\langle\frac{1}{V^2}\sum_{j}\frac{1}{\mu_j^2}\right\rangle \leq \frac{24}{Vm^2} \quad\stackrel{\longrightarrow}{_{V\rightarrow \infty}} \, 0.
\label{Flavour-Dies}
\end{equation}
The summatory is removed, adding a factor equal to the number of eigenvalues $24V$. This result states that \emph{there is no Aoki phase in lattice QCD with Ginsparg-Wilson fermions}. But we do not know yet whether the Lagrangian symmetries are spontaneously broken or not. Let's look at the next order parameter

\begin{equation}
\langle\left(i\bar\psi\gamma_5\psi\right)^2\rangle = 2\left\langle\frac{1}{V^2}\sum_{j}\frac{1}{\mu_j^2}\right\rangle - 4\left\langle\left(\frac{1}{V}\sum_{j}\frac{1}{\mu_j}\right)^2\right\rangle.
\label{Par-Dies}
\end{equation}
The first term of the r.h.s. is just equal to (\ref{Flavour-Dies}), so it must vanish in the thermodynamic limit. The second term is minus the square of a real quantity, then it must be negative or zero. The requirement (which we will assume) that $i\bar\psi\gamma_5\psi$ be an hermitian operator sets to zero this second term in the thermodynamic limit, for the expectation value of the square of an hermitian operator must be positive, thence

\begin{equation}
\lim_{V\rightarrow\infty}\left\langle\left(\frac{1}{V}\sum_{j}\frac{1}{\mu_j}\right)^2\right\rangle = 0.
\label{Par-Dies2}
\end{equation}
As both terms in the r.h.s. of (\ref{Par-Dies}) go to zero as the volume increases, this proves that parity is not spontaneously broken in lattice QCD with two flavours of Ginsparg-Wilson fermions\footnote{The result can be extended to any number of flavours with not much effort.}, at least for one of the more standard order parameters. In fact, we know that there exists an index theorem for Ginsparg-Wilson fermions \cite{VLaliena}, thus we can relate the zero modes of $D$ to the topological charge density,

\begin{equation}
\left\langle\left(\frac{1}{V}\sum_{j}\frac{1}{\mu_j}\right)^2\right\rangle = \left\langle\left(\frac{Q}{mV}\right)^2\right\rangle\left(\frac{2}{2+am}\right)^2 = -\frac{\chi_T}{Vm^2}\left(\frac{2}{2+am}\right)^2.
\label{Index}
\end{equation}
Taking into account (\ref{Par-Dies2}), we deduce that the topological charge density distribution function must be a Dirac delta centered in the origin.

Since we proved that both terms in the r.h.s. of (\ref{Par-Dies}) must vanish independently in the thermodynamic limit, this result also applies to a single flavoured condensate

\begin{equation}
\langle\left(i\bar\psi_u\gamma_5\psi_u\right)^2\rangle = \left\langle\frac{1}{V^2}\sum_{j}\frac{1}{\mu_j^2}\right\rangle - \left\langle\left(\frac{1}{V}\sum_{j}\frac{1}{\mu_j}\right)^2\right\rangle,
\label{Par1Flv}
\end{equation}
and, by extension, to any linear combination of the single-flavoured condensates $\bar\psi_j \gamma_5 \psi_j$.

As far as flavour symmetry is concerned, we have proved that $\left\langle\left(i\bar\psi\gamma_5\tau_3\psi\right)^2\right\rangle$ vanish in the infinite volume limit, but this is not enough, as this expectation value is forced to be zero because of parity conservation. Thus, we would like to investigate the quantity

\begin{equation}
\left|\langle\left(\bar\psi\tau_3\psi\right)^2\rangle\right| = \frac{2}{V^2}\left|\left\langle\sum_{j}\frac{1}{\left(\bar\lambda_j+m\right)^2}\right\rangle\right| \leq \frac{2}{V^2}\left\langle\sum_{j}\frac{1}{\left|\left(\bar\lambda_j+m\right)^2\right|}\right\rangle =$$
$$\frac{2}{V^2}\left\langle\sum_{j}\frac{1}{\left[\left[Re\left(\bar\lambda_j\right)+m\right]^2 + Im^2\left(\bar\lambda_j\right)\right]^2}\right\rangle \leq \frac{24}{Vm^2} \stackrel{\longrightarrow}{_{V\rightarrow \infty}} \, 0.
\label{Flavour-Dies2}
\end{equation}
Thus we can affirm that neither parity nor flavour are spontaneously broken in this regularization. At zero mass, we cannot establish an upper bound for the observables, hence the argument is not valid anymore. The fundamental question is: Why can we bound from above the value of these obervables at non-zero fermion mass?. The answer is related to a property of the Ginsparg-Wilson operator, that is, $\left\{D^{-1},\gamma_5\right\} = aR\gamma^5$, with $R$ a \emph{local} operator. We can write a similar equation for Wilson fermions, where $R$ is a \emph{non-local} operator, but in the case of Ginsparg-Wilson fermions, the locality of $R$ make the eigenvectors of $D$ look like chiral solutions at long distances. So, no quasi-chiral, exceptional configurations, at non-zero mass, are allowed, the Aoki phase is completely forbidden, and therefore, the symmetries are respected.

Other interesting results are straightforward from this point on. For instance, we can relate the transverse suceptibility, the topological susceptibility and the chiral condensate. First of all, we compute the transverse susceptibility,

\begin{equation}
\chi_5 = V\left\langle\left(i\bar\psi\gamma_5\psi\right)^2\right\rangle = \left\langle\frac{2}{V}\sum_{j}\frac{1}{\mu_j^2}\right\rangle + \frac{4\chi_T}{m^2}\left(\frac{2}{2+am}\right)^2.
\label{Trans-I}
\end{equation}
Now we write the first summand of the r.h.s. in terms of the chiral condensate, by making use of $\det{A} = \det{\gamma_5 A}$:

\begin{equation}
\left\langle\bar\psi\psi\right\rangle = -\left\langle\frac{1}{V}\frac{d}{dm}\ln\det\left(D+m\right)^{2}\right\rangle = -\frac{2}{V}\left\langle\sum_j\frac{m}{\mu_j^2}\right\rangle + O(a) + O(ma^2),
\label{ChCon}
\end{equation}
so we arrive at

\begin{equation}
\chi_5 = -\frac{\left\langle\bar\psi\psi\right\rangle}{m} + \frac{4\chi_T}{m^2},
\label{Trans-II}
\end{equation}
where we have dropped the factor $\left(\frac{2}{2+am}\right)^2$ assuming that we are close to the continuum limit. This relationship is not new at all, what we are showing here is simply a way to derive it. The interesting conclusion comes taking the vanishing mass limit $m\rightarrow 0$. Then, as the $\eta$ is not a Goldstone boson, $\chi_5$ must remain finite, so

\begin{equation}
\lim_{m\rightarrow 0} m\chi_5 = 0 \Rightarrow \lim_{m\rightarrow 0} \chi_T = \frac{m}{4}\left\langle\bar\psi\psi\right\rangle \propto f_\pi^2 m_\pi^2.
\label{Trans-F}
\end{equation}

\section{Final remarks}
The p.d.f. formalism can be used to cast some light on the old aim of understanding the realization of symmetries of QCD from first principles. Applying the p.d.f. to the Wilson regularization, we can explore certain, somewhat overlooked, properties of the Aoki phase. In fact, the p.d.f. states that, either the fermionic bilinear $i\bar\psi\gamma_5\psi$ can take non-zero values in the Aoki phase, extending thus the current picture of the phase diagram, or there exists an infinite tower of sum-rules the eigenvalues of the Dirac-Wilson operator must comply with. So far, no theoretical argument is strong enough to prove one of these scenarios to be right, thus a dynamical fermion simulation is mandatory at this point.

But the most interesting conclusions appear when we apply the p.d.f. formalism to the Ginsparg-Wilson regularization. There, we see how parity and vector-like symmetries must be realized for a non-vanishing fermion mass. This is a major result that overcomes the difficulties found by \cite{Vafa1,Vafa2}.

\acknowledgments

This work has been partially supported by an INFN-MEC collaboration, CICYT (grant FPA2006-02315) and DGIID-DGA (grant2007-E24/2). E. Follana is supported by Ministerio de Ciencia e Innovaci\'on through the Ram\'on y Cajal program.


\begin{thebibliography}{99}
\bibitem{Vafa1} 
C. Vafa and E. Witten, 
\textit{Phys. Rev. Lett.} \textbf{53}, (1984) 535. 

\bibitem{Vafa2} 
C. Vafa and E. Witten, 
\textit{Nucl. Phys. B} \textbf{234}, (1984) 173. 

\bibitem{agVafa} 
V. Azcoiti and A. Galante, 
\textit{Phys. Rev. Lett.} \textbf{83}, (1999) 1518;
X. Ji, 
\textit{Phys. Lett. B} \textbf{554}, (2003) 33;
P. R. Cromptom, 
\textit{Phys. Rev. D} \textbf{72}, (2005) 076003. 

\bibitem{yo}
V. Azcoiti, G. di Carlo and A. Vaquero,
\textit{JHEP} \textbf{0804}, (2008) 035 [{\tt arXiv:0804.1338}].

\bibitem{Aoki}
S. Aoki, 
\textit{Phys. Rev. D} \textbf{30}, (1984) 2653; 

\bibitem{Aoki2}
S. Aoki, 
\textit{Phys. Rev. Lett.} \textbf{57}, (1986) 3136. 

\bibitem{pdf}
V. Azcoiti, V. Laliena and X.Q. Luo, 
\textit{Phys. Lett. B} \textbf{354}, (1995) 111. 

\bibitem{VicSU2}
R. Aloisio, V. Azcoiti, G. di Carlo, A. Galante and A. F. Grillo,
\textit{Nucl. Phys. B} \textbf{606}, (2001) 322 [{\tt hep-lat/0011079}].

\bibitem{yo2}
V. Azcoiti, G. di Carlo and A. Vaquero,
\textit{Phys. Rev. D} \textbf{79}, (2009) 014509 [{\tt arXiv:0809.2972}].

\bibitem{Sharpe2}
S. Sharpe and R. Singleton, Jr.,
\textit{Phys. Rev. D} \textbf{58}, (1998) 074501 [{\tt hep-lat/9804028}].

\bibitem{LeutSmilg}
H. Leutwyler and A. Smilga,
\textit{Phys. Rev. D} \textbf{46}, (1992) 5607.

\bibitem{Sharpe}
S. Sharpe,
\textit{Phys. Rev. D} \textbf{79}, (2009) 054503 [{\tt arXiv:0811.0409}].

\bibitem{Niedermayer}
Ferenc Niedermayer,
\textit{Nucl. Phys. Proc. Suppl.} \textbf{73}, (1999) 105-119 [{\tt hep-lat/9810026}].

\bibitem{VLaliena}
P. Hasenfratz, V. Laliena and F. Niedermayer,
\textit{Phys. Lett. B} \textbf{427}, (1998) 125 [{\tt hep-lat/9801021}].

\end{thebibliography}
\end{document}